\documentclass{aa}
\usepackage{graphicx}
\usepackage[longnamesfirst]{natbib}
\usepackage{txfonts}
\begin{document}
\shortcites{baraffe1998}\shortcites{baraffe2003}\shortcites{baraffe2004}\shortcites{bakos2006}
\shortcites{bakos2007}\shortcites{bonavita2007}\shortcites{chauvin2007}
\shortcites{cutri2003}\shortcites{cushing2005}\shortcites{desidera2007}\shortcites{eggenberger2006}
\shortcites{eggenberger2007}\shortcites{fischer2007}\shortcites{holmberg2007}\shortcites{lagrange2006}
\shortcites{luhman2002}\shortcites{luhman2007}\shortcites{liu2007}\shortcites{locurto2006}
\shortcites{mayor2004}\shortcites{monet2003}\shortcites{mugrauer2007a}\shortcites{neuhaeuser2007}
\shortcites{mugrauer2004a}\shortcites{mugrauer2005b}\shortcites{mugrauer2004b}
\shortcites{mugrauer2006b}\shortcites{mugrauer2007b}\shortcites{patience2002}
\shortcites{raghavan2006}\shortcites{reid2004} \shortcites{skrutskie2006}\shortcites{tamuz2008}

\title{The multiplicity of exoplanet host stars \thanks{Based on observations obtained
on La Silla in ESO programs 079.C-0099(A), 080.C-0312(A)}}

\subtitle{New low-mass stellar companions of the exoplanet host stars\linebreak HD\,125612 and
HD\,212301}

\author{M. Mugrauer \inst{1} \and R. Neuh\"auser \inst{1}}

\offprints{Markus Mugrauer, markus@astro.uni-jena.de}

\institute{Astrophysikalisches Institut und Universit\"{a}ts-Sternwarte Jena, Schillerg\"a{\ss}chen 2-3,
07745 Jena, Germany}

\date{A\&A in press}

\abstract {}{We present new results from our ongoing multiplicity study of exoplanet host stars,
carried out with SofI/NTT. We provide the most recent list of confirmed binary and triple star
systems that harbor exoplanets.}{We use direct imaging to identify wide stellar and substellar
companions as co-moving objects to the observed exoplanet host stars, whose masses and spectral
types are determined with follow-up photometry and spectroscopy.}{We found two new co-moving
companions of the exoplanet host stars HD\,125612 and HD\,212301. HD\,125612\,B is a wide M4 dwarf
(0.18\,$M_\odot$) companion of the exoplanet host star HD\,125612, located about 1.5\,arcmin
($\sim$4750\,AU of projected separation) south-east of its primary. In contrast, HD\,212301\,B is a
close M3 dwarf (0.35\,$M_\odot$), which is found about 4.4\,arcsec ($\sim$230\,AU of projected
separation) north-west of its primary.}{The binaries HD\,125612\,AB and HD\,212301\,AB are new
members in the continuously growing list of exoplanet host star systems of which 43 are presently
known. Hence, the multiplicity rate of exoplanet host stars is about 17\,\%.}


\maketitle

\section{Introduction}

For more than a decade, radial-velocity and photometric planet search campaigns have indirectly
identified more than three hundred exoplanets that revolve around mostly sun-like stars, in the
solar neighborhood. The majority of these exoplanet host stars are isolated single stars, but
during recent years more and more of them turned out to be the brighter primary component of
stellar systems, identified by ongoing multiplicity studies. These studies were carried out with
seeing limited near infrared imaging \cite[see e.g.][2004b, 2005b, 2006a, 2007a,
2007b]{mugrauer2004a}\nocite{mugrauer2004b}\nocite{mugrauer2005b}\nocite{mugrauer2006a}\nocite{mugrauer2007a}\nocite{mugrauer2007b},
high contrast diffraction limited AO observations \citep[see e.g. Patience et al. (2002), Luhman \&
Jayawardhana (2002), Chauvin et al. 2006,][and most recently Eggenberger et al.
2007]{neuhaeuser2007}\nocite{eggenberger2007}\nocite{chauvin2006}\nocite{patience2002}\nocite{luhman2002}\nocite{neuhaeuser2007},
as well as from space \citep{luhman2007}. In addition, data from visible and infrared all sky
surveys like POSS or 2MASS are used to identify new companions of exoplanet host stars \citep[as
reported e.g. by Bakos et al. 2006, Raghavan et al. 2006, or most recently by][]{desidera2007}
\nocite{raghavan2006}\nocite{bakos2006}.

Most of the detected stellar companions of exoplanet host stars are low-mass main sequence stars
with projected separations of between a few tens up to more than 10000\,AU. In a few cases the
companions themselves turned out to be close binaries, i.e. these systems are hierarchical triples
\citep[see][for a summary]{mugrauer2007a}. The closest of these systems presently known is
HD\,65216\,A+BC, with a projected separation of about 250\,AU \citep{mugrauer2007b}. Also white
dwarfs have been identified as companions of exoplanet host stars, suggesting that exoplanets can
survive the post main sequence evolution of a nearby star \citep[e.g. Gl\,86\,B, $\sim$20\,AU, see
][for more details]{mugrauer2005a}. Later on, the first directly imaged substellar companion of an
exoplanet host star, the T7-T8 brown dwarf HD\,3651\,B, was discovered \citep[see Mugrauer et al.
2006b, and][]{liu2007}\nocite{mugrauer2006b}. In addition to these imaging surveys a dynamical
characterization of the closest exoplanet host binaries, like Gl\,86\,AB
\citep[see][]{lagrange2006}, or $\gamma$\,Cep\,AB \cite[see][]{neuhaeuser2007}, is being carried
out to determine the full set of their orbital elements.

All these efforts will help to reveal the true impact of stellar multiplicity on the formation
process of planets and the evolution of their orbits. For recent statistical studies we refer here
to e.g. \cite{mugrauer2007c} or \cite{bonavita2007}.

In this paper we present new results of our ongoing multiplicity study carried out at La Silla
observatory with SofI/NTT. We detected two new low-mass stellar companions of the exoplanet host
stars HD\,125612 and HD\,212301. Our SofI astro- and photometry is described in section\,2, the
follow-up spectroscopy in section\,3. The properties of the newly found exoplanet host binaries are
described in section\,4, where the SofI detection limits also are presented and the separation
space of possible, so far undetected, additional companions is discussed. A list of all presently
known and confirmed exoplanet host star systems is given in the Appendix.

\section{Astro- and photometry}

\subsection{HD\,125612}

The exoplanet host star HD\,125612 is a solar-like G3V star \citep[$1.1\pm0.07\,M_{\odot}$,
][]{fischer2007}, which can be found on the sky between the constellations Virgo and Libra at a
distance of $53\pm$4\,pc \citep[Hipparcos, ][]{perryman1997}. According to \cite{fischer2007}
HD\,125612 shows only weak chromospheric activity ($log R^{'}_{\rm HK}=-4.85$), and its
chromospheric age of $3.3\pm2$\,Gyr is comparable with the age determined with evolutionary models
(0.16 to 5.6\,Gyr). In our ongoing multiplicity study of exoplanet host stars, HD\,125612 was
observed the first time in June 2007 with SofI/NTT. All our SofI images are astrometrically
calibrated with the 2MASS point source catalogue \citep{skrutskie2006}, as summarized in
Table\,\ref{table_astrocal}.

\begin{table}[htb] \caption{The astrometrical calibration of SofI/NTT.
The pixel scale $PS$ and the detector position angle $PA$ with their uncertainties are listed. The
detector is tilted by $PA$ from north to west.}
\begin{center}
\begin{tabular}{c|c|c|c}
\hline\hline
instrument & epoch & $PS$ [$''$] & $PA$ [$^{\circ}$]\\
\hline
SofI$_{large}$ & 06/2007 & 0.28791$\pm$0.00021 & 89.984$\pm$0.026\\
SofI$_{large}$ & 01/2008 & 0.28794$\pm$0.00025 & 89.997$\pm$0.056\\
\hline\hline
\end{tabular}
\label{table_astrocal}
\end{center}
\end{table}

\begin{figure} [htb]
\resizebox{\hsize}{!}{\includegraphics{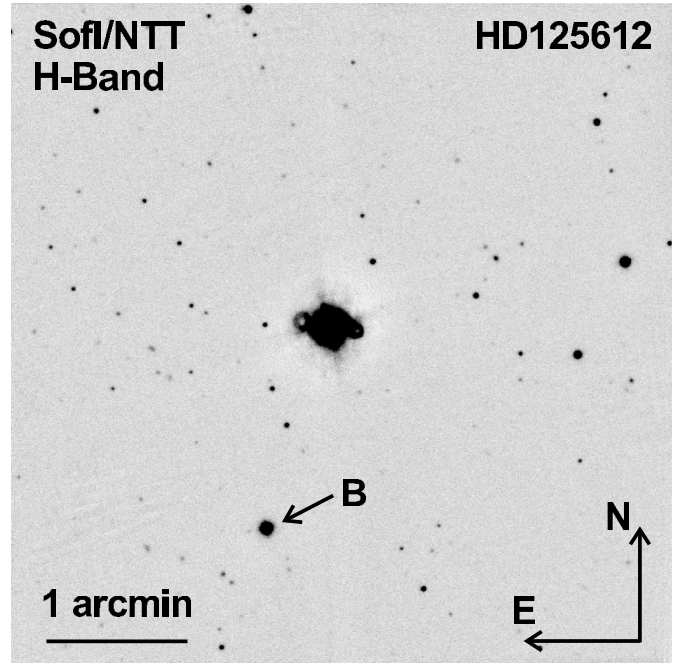}}\caption{SofI large field image of the
exoplanet host star HD\,125612. The detected co-moving companion of the exoplanet host star is
indicated with a black arrow.}\label{pic_hd125}
\end{figure}

Our reduced SofI H-band image of HD\,125612 is shown in Fig.\,\ref{pic_hd125}. The total
integration time of this image is 10\,min taken with the standard set-up of our programme
\citep[for further details see e.g.][]{mugrauer2007a}. Several faint companion candidates down to
H$\sim$18\,mag ($S/N=10$) are detected around the exoplanet host star, which is located in the
center of the image.

\begin{figure} [htb]
\resizebox{\hsize}{!}{\includegraphics{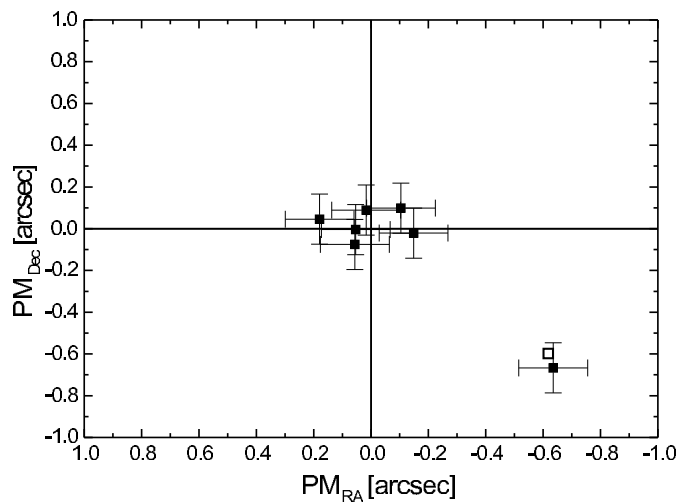}}\caption{The derived motion of all companion
candidates detected in our SofI H-band image and also imaged by 2MASS around the exoplanet host
star. The proper and parallactic motion of HD\,125612 for the give epoch difference is indicated
with a black square.}\label{pm_hd125}
\end{figure}

HD\,125612 was imaged by 2MASS in April 1998, i.e more than 9 years before our SofI 1st epoch
observation. HD\,125612 exhibits a high proper motion ($\mu_{\rm RA}=-64.47\pm1.57$\,mas/yr
$\mu_{\rm Dec}=-65.64\pm1.03$\,mas/yr) which is well known from Hipparcos. By comparing the 2MASS
with our SofI image we derive the proper motion of all companion candidates detected in our SofI,
and also in the less sensitive 2MASS image (see Fig.\,\ref{pm_hd125}).

The majority of detected objects only exhibits small or negligible proper motion. Hence, all these
candidates are unrelated background objects only randomly located close to the line of sight in the
direction of the exoplanet host star. In contrast, one candidate located about 89.98\,arcsec
($\sim$4750\,AU of projected separation) south-east of HD\,125612 clearly shares the proper motion
of HD\,125612. This newly found co-moving companion, indicated with a black arrow in
Fig.\,\ref{pic_hd125}, will be denoted as HD\,125612\,B and the exoplanet host star as
HD\,125612\,A, from here on.

We also measure the separation and the position angle of HD\,125612\,B relative to its primary in
the 2MASS, as well as in our SofI H-band image. This relative astrometry of the companion is
summarized in Table\,\ref{table_seppa}. Neither separation nor position angle of HD\,125612\,B
change significantly during the more than 9\,years of epoch difference, as is expected for a
co-moving companion.

\begin{table}[htb] \caption{The separations and position angles of HD\,125612\,B and HD\,212301\,B relative
to their primaries --- the exoplanet host stars HD\,125612\,A, and HD\,212301\,A. In the columns
sep$_{~if~bg}$ and PA$_{~if~bg}$ we show the expected separation and position angle in the case
that both companions are non-moving background objects.}
\begin{center}
\begin{tabular}{c|c|c|c}
\hline\hline
& epoch & sep$_{~obs}$[arcsec] & sep$_{~if~bg}$ [arcsec]\\
\hline
HD\,125612\,B & 2MA 04/98 & 89.957$\pm$0.055 & 89.957$\pm$0.055\\
              & NTT 06/07 & 89.994$\pm$0.066 & 89.574$\pm$0.056\\
\hline
HD\,212301\,B & 2MA 12/99 &   4.34$\pm$0.23  & 4.34$\pm$0.23\\
              & NTT 06/07 &   4.36$\pm$0.06  & 5.05$\pm$0.23\\
              & NTT 01/08 &   4.43$\pm$0.06  & 5.08$\pm$0.23\\
\hline\hline
& epoch & PA [$^{\circ}$]   & PA$_{~if~bg}$ [$^{\circ}$]\\
HD\,125612\,B & 2MA 04/98 & 162.696$\pm$0.070 & 162.696$\pm$0.070\\
              & NTT 06/07 & 162.682$\pm$0.052 & 162.205$\pm$0.071\\
\hline
HD\,212301\,B & 2MA 12/99 & 275.7$\pm$3.0 & 275.7$\pm$3.0\\
              & NTT 06/07 & 275.1$\pm$0.5 & 283.0$\pm$3.0\\
              & NTT 01/08 & 275.8$\pm$0.5 & 283.3$\pm$3.0\\
\hline\hline
\end{tabular}
\end{center}\label{table_seppa}
\end{table}

The proper motion of HD\,125612\,B is also listed in the USNO-B1.0 catalogue \citep{monet2003}
($\mu_{\rm RA}=-64\pm1$\,mas/yr, $\mu_{\rm Dec}=-58\pm4$\,mas/yr) and is fully consistent with the
Hipparcos proper motion of the exoplanet host star. This is further proof of the astrometric
companionship of HD\,125612\,B to the exoplanet host star.

We determine the H-band photometry of HD\,125612\,B in our SofI image and obtain
H=11.761$\pm$0.034\,mag, which is fully consistent with its 2MASS photometry
H=11.773$\pm$0.023\,mag. In addition, the 2MASS point source catalogue also lists the J-, and
$K_{\rm S}$-band photometry of HD\,125612\,B ($J=12.381\pm0.0244$\,mag, and $K_{\rm
S}=11.514\pm0.024$\,mag). The well-known Hipparcos distance of the exoplanet host star, and the
given 2MASS apparent magnitudes of HD\,125612\,B, finally yield the absolute photometry of the
companion: $M_{J}=8.77\pm0.16$\,mag, $M_{H}=8.16\pm0.16$\,mag, $M_{K_{s}}=7.90\pm0.16$\,mag.

According to the evolutionary models for low-mass stars from \cite{baraffe1998} the derived
absolute photometry of HD\,125612\,B is consistent with a $0.184\pm0.012$\,$M_{\odot}$ star for an
assumed age between 1 and 5\,Gyr. According to the magnitude-spectral type relation from
\cite{reid2004}, the absolute photometry of HD\,125612\,B is consistent with an M4 dwarf. The
spectral type estimation has to be confirmed by follow-up spectroscopy (see next section).

\subsection{HD\,212301}

The exoplanet host star HD\,212301 is a 1.9 to 5.4\,Gyr \citep{holmberg2007} old F8 dwarf
\citep[$1.27\pm0.02\,M_{\odot}$, ][]{locurto2006}, which is located at a distance of $53\pm2$\,pc,
\citep[Hipparcos, ][]{perryman1997} in the constellation Octans.

\begin{figure} [htb]
\resizebox{\hsize}{!}{\includegraphics{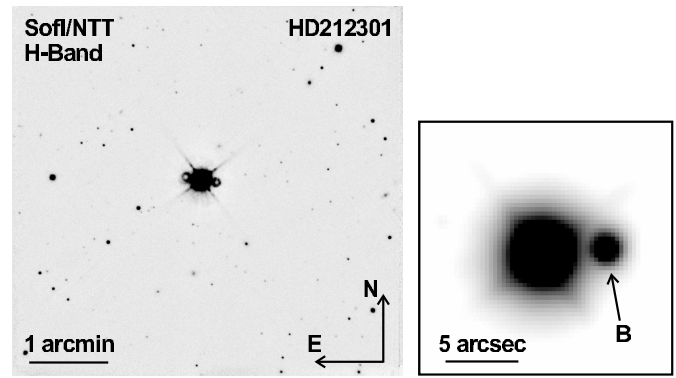}}\caption{\textbf{Left:} Our 1st epoch SofI
H-band image of the exoplanet host star HD\,212301. \textbf{Right:} Detail of the whole SofI image,
showing the central region around the exoplanet host star, using a logarithmic scaling. A close
companion candidate is detected only 4\,arcsec north-west of HD\,212301, marked with a black
arrow.}\label{pic_hd212}
\end{figure}

We observed HD\,212301 twice with SofI in H-band, with a total integration time of 10\,min in June
2007 as well as in January 2008. Our 1st epoch SofI image is shown in Fig.\,\ref{pic_hd212}.
Several faint companion candidates, down to H=18\,mag ($S/N=10$), are detected around the bright
exoplanet host star. We could also identify a close candidate, located in both SofI images only
about 4.38\,arcsec ($\sim$\,230\,AU of projected separation) north-west of the HD\,212301.

In order to check if this close candidate is also detected by 2MASS, we carefully inspected the
2MASS J-, H-, and K$_{\rm S}$-band images. We found that in all 2MASS images the PSF of HD\,212301
appears elongated in the direction where the close candidate is located. In contrast, the PSF of
objects found around the star in the 2MASS images all exhibit radially symmetric PSFs. We
deconvolved all 2MASS images, using the PSFs of objects detected around HD\,212301 as a reference.
The object functions of the exoplanet host star in the J- and K$_{\rm S}$-band remain elongated,
while in the H-band the two separated object functions of HD\,212301 and its close companion
candidate could be reconstructed.

\begin{figure} [htb]
\resizebox{\hsize}{!}{\includegraphics{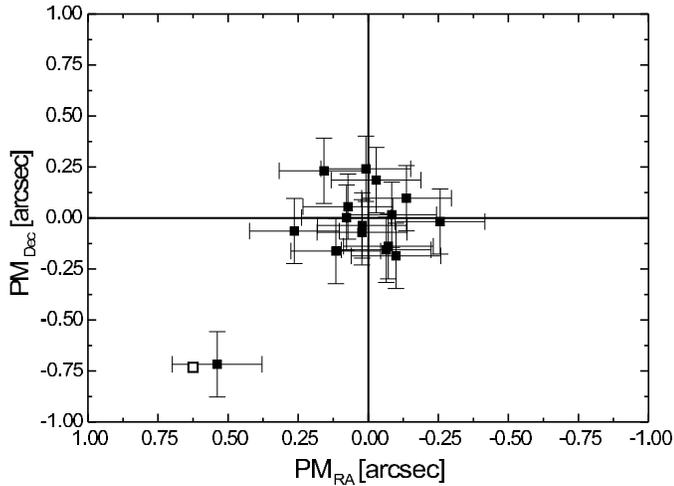}}\caption{The derived proper motion of all
companion candidates detected around the exoplanet host star HD\,212301 in our 2nd epoch SofI, and
the deconvolved 2MASS H-band image. The proper and parallactic motion of the exoplanet host star
for the give epoch difference is indicated with a black square.}\label{pm_hd212}
\end{figure}

By comparing our two SofI images of HD\,212301 with the deconvolved 2MASS H-band image taken at the
end of 1999, we can determine the proper motion of all detected objects. Due to the high proper
motion of HD\,212301 ($\mu_{\rm RA}=76.11\pm0.85$\,mas/yr and $\mu_{\rm
Dec}=-91.64\pm0.63$\,mas/yr, from Hipparcos) and the long epoch difference of more than seven years
between the 2MASS and our SofI images, real companions of HD\,212301 can easily be identified as
co-moving objects. The derived proper motions of all candidates between the 2MASS and our 2nd epoch
SofI observation are shown in Fig.\,\ref{pm_hd212}.

Only the close companion candidate clearly shares the proper motion of the exoplanet host star.
Hence, this is a new co-moving companion of HD\,212301, called HD\,212301\,B from here on.

Beside the proper motion of HD\,212301\,B we also determined its relative astrometry to its primary
in all observing epochs, summarized in Table\,\ref{table_seppa}. The separation and position angle
of HD\,212301\,B do not change significantly over time, as expected for a co-moving companion.

Because of its small angular separation HD\,212301\,B is not resolved in the 2MASS J, and K$_{\rm
S}$-band images. Therefore, we obtained additional photometry with SofI in June 2007 and observed
HD\,212301\,B in J, and K$_{\rm S}$ broad, as well as in the Br$_{\gamma}$ narrow band filter.
Within a jitter box of 20\,arcsecs we obtained three images in J and K$_{\rm S}$ and five images in
Br$_{\gamma}$ which are each the average of 50 exposures with an integration time of 1.2\,s, and
obtained $J=11.056\pm0.086$\,mag, and $K_{s}=10.210\pm0.058$\,mag. In the H-band we obtained
$H=10.516\pm0.041$\,mag in our 1st, and $H=10.587\pm0.053$\,mag in the 2nd SofI imaging epoch. The
K$_{\rm S}$ band magnitude of the exoplanet host star is listed in the 2MASS point source catalogue
(K$_{\rm S}=6.466\pm0.021$\,mag). Together with our SofI K$_{\rm S}$-band photometry of
HD\,212301\,B this yields a magnitude difference between HD\,212301\,A and B of $\Delta
\rm{K}_{\rm{S}} = 3.744\pm0.061$\,mag. This result is confirmed by the magnitude difference between
HD\,212301\,B and the exoplanet host star $\Delta \rm{Br}_{\gamma} = 3.67\pm0.06$\,mag, as measured
in our SofI Br$\gamma$ image. In addition, we also determined the photometry of HD\,212301\,B in
the deconvolved 2MASS H-band image and obtained H$_{\rm 2Mass}=10.71\pm0.17$\,mag which is
consistent with our SofI H-band photometry of the co-moving companion.

With the measured apparent photometry of HD\,212301\,B, and the well known distance of the
exoplanet host star we derive its absolute photometry, which is $M_{J}=7.45\pm0.12$,
$M_{H}=6.94\pm0.09$, and $M_{K_{s}}=6.60\pm0.10$.

With the evolutionary models of \cite{baraffe1998} and the derived absolute photometry of
HD\,212301\,B we determine the mass of the companion to be 0.35$\pm$0.02\,$M_{\odot}$, for an
assumed age of between 0.5 and 5\,Gyr. According to the magnitude-spectral type relation from
\cite{reid2004} we expect the spectral type of the companion to be M3V, which has to be confirmed
with follow-up spectroscopy (see the following section).

\section{Spectroscopy}

In June 2007, we obtained H- and K-band follow-up spectroscopy of the two co-moving companions
HD\,125612\,B and HD\,212301\,B with SofI. We used the grism \textsl{RED} in combination with a
1\,arcsec slit which offers a resolving power $\lambda / \Delta \lambda = 588$ and dispersion of
10.22\,\AA\,\,per pixel. For both co-moving companions we always took 10 frames each with an
integration time of 60\,s. Between individual exposures a nodding of 45\,arcsec between two
positions along the slit, as well as a 5\,arcsec random jitter was applied. For wavelength
calibration we took spectra of a Xenon lamp. Standard IRAF routines for spectroscopy were used for
data reduction. Telluric features in the reduced spectra were removed by dividing with spectra of
telluric standard stars, whose spectra were always taken directly after the spectroscopy of the
companions, with airmass difference between science and calibration spectra of less than 0.1. We
took spectra of HIP\,73881 (B2-3V) in the case of HD\,125612\,B, and HIP\,100170 (B2V) for
HIP\,212301, respectively. The spectral response function of SofI was determined using the spectra
of the telluric standards, as well as flux-calibrated spectra from the spectral library of
\cite{pickles1998}.

\begin{figure} [htb]
\resizebox{\hsize}{!}{\includegraphics{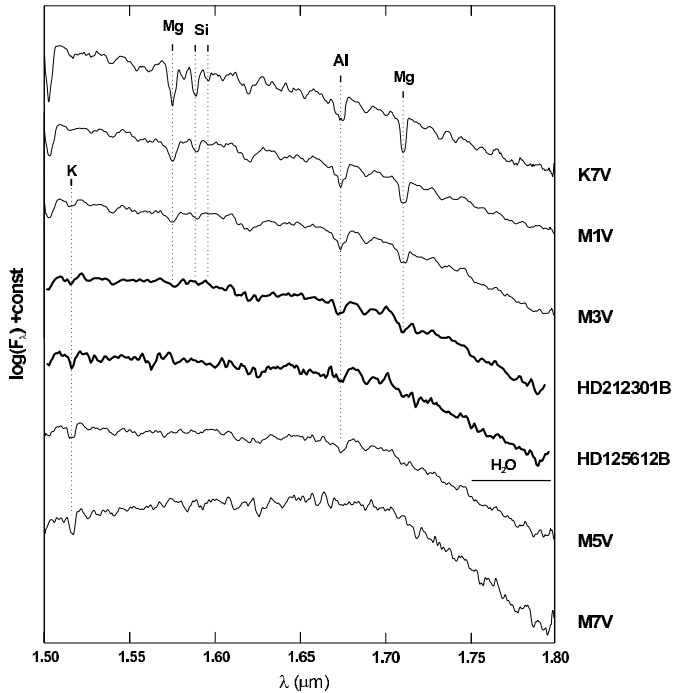}}\caption{The H-band SofI spectra of
HD\,125612\,B and HD\,212301\,B together with comparison spectra of dwarfs from the IRTF Spectral
Library \citep{cushing2005}. The most prominent spectral atomic and molecular features are
indicated.}\label{h_spec}
\end{figure}

\begin{figure} [htb]
\resizebox{\hsize}{!}{\includegraphics{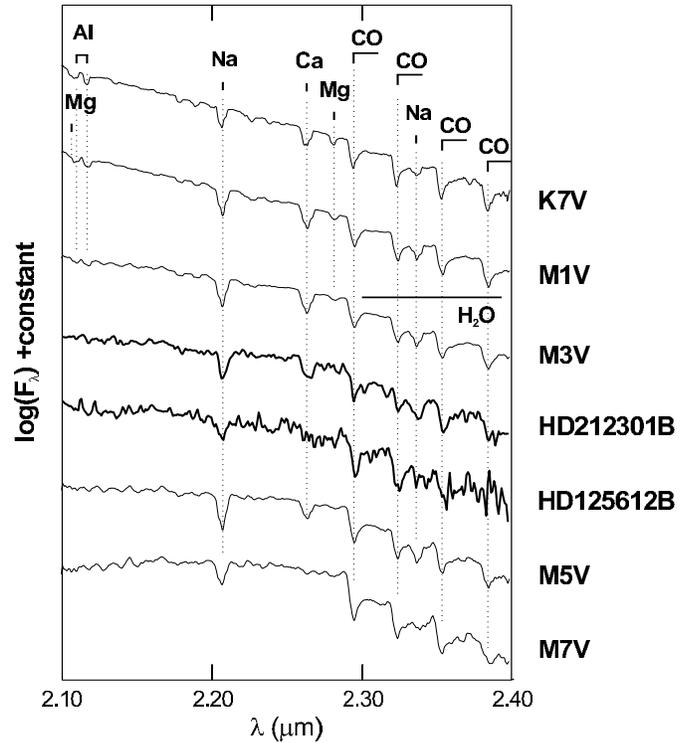}}\caption{The K-band SofI spectra of
HD\,125612\,B and HD\,212301\,B together with comparison spectra of dwarfs from the IRTF spectral
library (\cite{cushing2005}). The most prominent spectral atomic and molecular features are
indicated.}\label{k_spec}
\end{figure}

The flux-calibrated H- and K-band spectra of HD\,125612\,B and HD\,212301\,B are shown in
Fig.\,\ref{h_spec} and Fig.\,\ref{k_spec}. The spectra of both companions are compared with
template spectra from the IRTF spectral library \citep{cushing2005}, smoothed to the same
resolution as our SofI spectra ($\Delta \lambda / \lambda = 1/588$).

In the H-band spectrum of HD\,212301\,B the most prominent features are those of aluminium at
1.674\,$\mu$m, magnesium at 1.711\,$\mu$m and weaker at 1.576\,$\mu$m, potassium at 1.517\,$\mu$m,
and the faintly detected silicon line at 1.589\,$\mu$m. All detected spectral features, as well as
the continuum shape of the spectrum compare well with the M3V template spectrum. In contrast, the
continuum of the spectrum of HD\,125612\,B is more consistent with that of an M5 dwarf. In addition
the potassium line in the spectrum of HD\,125612\,B is stronger than that of the spectrum of
HD\,212301\,B, which indicates a slightly later spectral type of between M3 and M5V.

In the K-band, the spectra of both companions show the prominent absorption line doublet of sodium
at 2.208\,$\mu$m, and the molecular absorption bands of CO at wavelengths longer than
2.294\,$\mu$m, typical of M3 to M5 dwarfs. In contrast, the calcium absorption line doublet at
2.265\,$\mu$m which is well detected in the spectrum of HD\,212301\,B cannot be identified in the
spectrum of HD\,125612\,B. Because all atomic absorption features weaken to later spectral types
this indicates that the spectrum of HD\,125612\,B, is slightly later than that of HD\,212301\,B
mostly consistent with a M4 to M5 dwarf.

According to our SofI spectroscopy we can conclude that HD\,125126\,B and HD\,212301\,B are both
mid M dwarfs (M3 to M5V). Our SofI spectroscopy fully confirms the spectral type estimation of both
companions obtained from their apparent photometry, assuming that they are dwarfs located at the
distances of the exoplanet host stars (M4V for HD\,125612\,B, and M3V for HD\,212301\,B). Hence,
the companionship of HD\,125612\,B and HD\,212301\,B, which was first revealed with astrometry
(common proper motion), is finally confirmed by photometry and spectroscopy.

\section{Discussion}

As described in detail in the last sections HD\,125612\,B is a wide M4 dwarf companion with a mass
of $\sim$0.18\,$M_{\odot}$ which is separated by about 4750\,AU from its primary, the exoplanet
host star HD\,125612\,A. In contrast, HD\,212301\,B is a close M3 dwarf (0.35\,$M_{\odot}$)
companion, located at a separation of about 230\,AU from its primary.

\begin{figure} [htb]
\resizebox{\hsize}{!}{\includegraphics{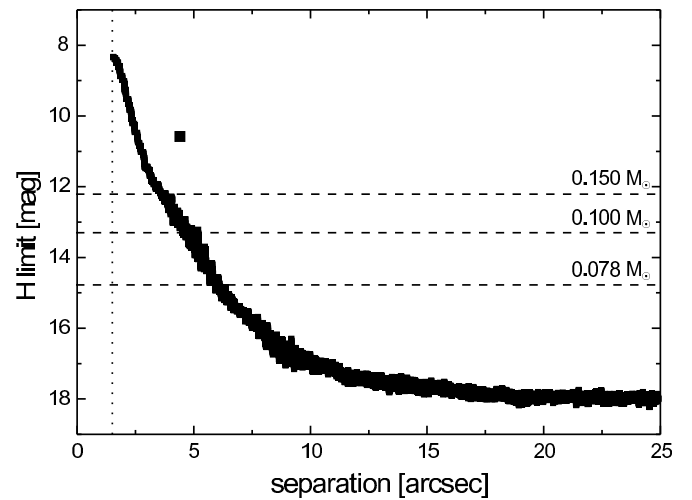}}\caption{The average detection limit
($S/N=10$) in our SofI images of the exoplanet host stars HD\,125612\,A and HD\,212301\,A plotted
for a range of angular separations up to 25\,arcsec. The detected co-moving companion HD\,212301\,B
is indicated as a black square.}\label{limit_hd212}
\end{figure}

The typical SofI detection limit for both exoplanet host stars is shown in Fig.\,\ref{limit_hd212}.
At an angular separation of less than 1.5\,arcsec ($\sim$80\,AU) saturation occurs, i.e. companions
cannot be detected in this region. According to the evolutionary models from \cite{baraffe1998} all
stellar companions ($mass>0.078\,M_{\rm Jup}$) are detectable beyond 6\,arcsec ($\sim$320\,AU)
around both stars, for assumed ages between 1 and 5\,Gyr. In the background limited region beyond
about 15\,arcsec ($\sim$ 800\,AU) a sensitivity of H=18\,mag ($S/N=10$) is reached, i.e. companions
with absolute magnitudes down to $M_{\rm H}=14.4$\,mag can be imaged in this region. This allows
the detection of brown dwarf companions with masses down to 37 to 65\,$M_{\rm Jup}$. Companions
with angular separations of up to about 143\,arcsec ($\sim$ 7500\,AU of projected separation) can
be detected together with the exoplanet host stars in our SofI images.

HD\,125612 was observed only in the 1st epoch with SofI. A 2nd epoch follow-up imaging could not be
carried out in 2008 due to bad weather. For HD\,212301 we obtained a 2nd epoch image in January
2008 but only to confirm the detection of the companion HD\,212301\,B and to rule out that it is a
fast moving foreground object e.g. a planetoid in the solar system. The expected total motion of
the companion between both observing epochs is only about 40\,mas, too small to be clearly detected
with SofI in its large field mode. Hence, follow up SofI imaging is needed for both stars with
sufficient epoch difference to check the companionship of all faint companion candidates not imaged
by 2MASS.

Nevertheless, we can already exclude additional companions around both exoplanet host stars based
on the 2MASS detection limit. According to \cite{skrutskie2006} the 2MASS H-band limit is 15.1\,mag
at $S/N=10$, the limit for all detected sources with accurate astrometry. As determined by us, this
limit is reached in the 2MASS images beyond about 20\,arcsec ($\sim$1060\,AU) around both stars.
According to the evolutionary models from \cite{baraffe1998} and an assumed age of 1 to 5\,Gyr
companions with masses down to 63 and 74\,$M_{\rm Jup}$ can be detected in 2MASS. All stellar
companions ($mass>0.078$\,$M_{\rm Jup}$) can be detected beyond about 13\,arcsec ($\sim$690\,AU)
around both stars.

In addition, we can also rule out companions that should not be on long-term stable orbits around
the exoplanet host stars in both binary systems. According to \cite{holman1999} only companions
with semi-major axes smaller than the critical semi-major axis exhibit long-term stable orbits. In
the case of the HD\,125612\,AB system (HD\,125612\,A ($\sim$1.1\,$M_{\odot}$), HD\,125612\,B
($\sim$0.18\,$M_{\odot}$), 4750\,AU of projected separation) we expect the long-term stable zone to
extend up to $\sim$650\,AU (12.4\,arcsec of angular separation), assuming that HD\,125612\,B is on
a circular orbit, whose semi-major axis corresponds to the observed projected separation of the
companion. This value can be considered as an upper separation limit for additional long-term
stable companions. In the case of an eccentric orbit of HD\,125612\,B the extent of this long-term
stable zone should be smaller. Indeed, there is one faint companion candidate in our SofI image
detected about 10\,arcsec west of HD\,125612\,A. Follow-up imaging is needed to test the
companionship of this faint companion candidate.

In the case of the HD\,212301\,AB system (primary mass of about 1.27\,$M_{\odot}$, secondary mass
of $\sim$\,0.35\,$M_{\odot}$, with a projected separation of $\sim$\,230\,AU) we derive an extent
of the long-term stable zone of only $\sim$\,38\,AU, i.e about 0.7\,arcsec of angular separation.
No objects are detected within this given angular radius around HD\,212301\,A in our SofI images.

The binaries HD\,125612\,AB and HD\,212301\,AB are two new members of the growing list of exoplanet
host multiple star systems. Today 250 exoplanet host stars are known and ongoing multiplicity
studies so far have found 43 of them to be components of a multiple star system (see the Appendix
for summary). Hence, the multiplicity-rate of the exoplanet host stars is at least 17\,\%.

\acknowledgements {We would like to thank the technical staff of the ESO NTT. We made use of the
2MASS public data releases as well as the Simbad database operated at the Observatoire Strasbourg.}

%\bibliography{ref}

\begin{appendix}

\section{Presently known exoplanet host star systems}

Currently, 43 confirmed star systems are known that harbor at least one exoplanet; 37 of them are
binaries, and the remaining 6 are triples. There is only one system known with a directly detected
wide substellar companion, namely HD\,3651\,AB \citep{mugrauer2006b}.

\subsection{Exoplanet host binary star systems}

\flushleft{55\,Cnc\,AB\footnote{For all exoplanet host star systems listed, the first component is
always the exoplanet host star, the second one is its stellar companion.}, 83\,Leo\,BA,
GJ\,3021\,AB, GJ\,777\,AB, Gl\,86\,AB, HD\,114762\,AB, HD\,142\,AB, HD\,195019\,AB,
$\upsilon$\,And\,AB, HD\,222582\,AB, HD\,27442\,AB, HD\,46375\,AB, $\tau$\,Boo\,AB, HD\,80606\,AB,
HD\,41004\,AB, HD\,19994\,AB, HD\,11964\,AB \cite[for details see e.g.][]{mugrauer2007a};
HD\,147513\,AB \citep{mayor2004}; HD\,75289\,AB \citep{mugrauer2004a}; HD\,89744\,AB
\citep{mugrauer2004b}; HD\,16141\,AB, HD\,114729\,AB, and HD\,213240\,AB \citep[all described in
detail in][]{mugrauer2005b}; HD\,38529\,AB, and HD\,188015\,AB \citep[both reported
by][]{raghavan2006}; HD\,189733\,AB \citep{bakos2006}; HD\,142022\,AB \citep{eggenberger2006};
HD\,196885\,AB \citep{chauvin2007}; HD\,20782\,AB, and HD\,109749\,AB \citep[both described
by][]{desidera2007}; $\gamma$\,Cep\,AB \citep[whose B component was directly detected first by][who
also clarified the true nature of this component]{neuhaeuser2007}; HD\,101903\,AB
\citep{mugrauer2007b}; HD\,177830\,AB \citep{eggenberger2007}; ADS\,16402\,BA \citep{bakos2007};
HD\,156846\,AB \citep{tamuz2008}; HD\,125612\,AB, and HD\,212301\,AB (both presented in this work)}

\subsection{Exoplanet host triple star systems}

\flushleft{HD\,178911\,B+AC, HD\,40979\,A+BC, 16\,Cyg\,B+AC, and HD\,219449\,A+BC \cite[see][for
details]{mugrauer2007a}; HD\,196050\,A+BC \citep[see Mugrauer et al. 2005b
and][]{eggenberger2007}\nocite{mugrauer2005b}; HD\,65216\,A+BC \citep{mugrauer2007b}}

\end{appendix}

\end{document}